\documentclass[11pt,twoside]{article}
\usepackage{asp2014}

\aspSuppressVolSlug
\resetcounters

\begin{document}

\title{Web SAMP and HTTPS: What to do?}

\author{M.~B.~Taylor}
\affil{H.~H.~Wills Physics Laboratory,
       University of Bristol, UK;
       \email{m.b.taylor@bristol.ac.uk}}
% remove/add as you need

% remove/add authors as you need
\paperauthor{M.~B.~Taylor}{m.b.taylor@bristol.ac.uk}{0000-0002-4209-1479}{University of Bristol}{School of Physics}{Bristol}{Bristol}{BS8 1TL}{U.K.}
% remove/add as you need

% leave these next few aindex lines commented for the editors to enable them. Use Aindex.py to generate them for yourself.
% first presenting author should be the first entry for bold-facing the author index page-reference
%\aindex{Taylor,~M.~B.}

% leave the ssindex lines commented for the editors to enable them, use Index.py to suggest yours
%\ssindex{protocols!SAMP}
%\ssindex{protocols!HTTPS}

% leave the ooindex lines commented for the editors to enable them, use ascl.py to suggest yours

\begin{abstract}

SAMP, the Simple Application Messaging Protocol, is a standard developed
within the Virtual Observatory to allow communication between different
software items on the desktop.  One popular usage scenario has been
enabling one-click transmission of a table or FITS image from a web page,
typically an archive search result of some kind, to a desktop application
such as TOPCAT, Aladin or ds9.  This has worked well for HTTP web pages
since the introduction of the SAMP Web Profile in SAMP 1.3, but the Web
Profile will not work over HTTPS, which is increasingly being adopted
by data providers.

This paper presents a summary of the problem and explores some possible
ways forward, for which working prototypes have been developed: specify
a new HTTPS-capable Profile, use a SAMP-capable helper application,
or abandon using SAMP over HTTPS.

\end{abstract}

\section{Introduction}

SAMP \citep{2015A&C....11...81T}
is middleware designed to allow loose interoperability
between astronomy applications on a user's desktop.
An example pattern of use is to send a catalogue from a
catalogue analysis tool such as TOPCAT to an image analysis
tool such as Aladin, so that activity in the two tools can
be linked: for instance the catalogue positions can be overplotted
on sky imagery, and if a user indicates a selection in a
colour-magnitude plot in the catalogue tool,
the corresponding objects can be highlighted in the image tool.

The architecture is based on message passing via a central {\em Hub\/},
a daemon that runs on the user's machine;
the hub may be either free-standing or embedded in
one of the running SAMP-aware applications.
Each SAMP client has to establish two-way communication with this hub,
which it does according to one of the {\em Profiles\/}
defined by the SAMP standard.
Initially, only the {\em Standard Profile\/}
was defined, in which clients locate the hub from a file in
the user's home directory and communicate via bi-directional
XML-RPC calls.

\begin{figure}
\includegraphics[width=\textwidth]{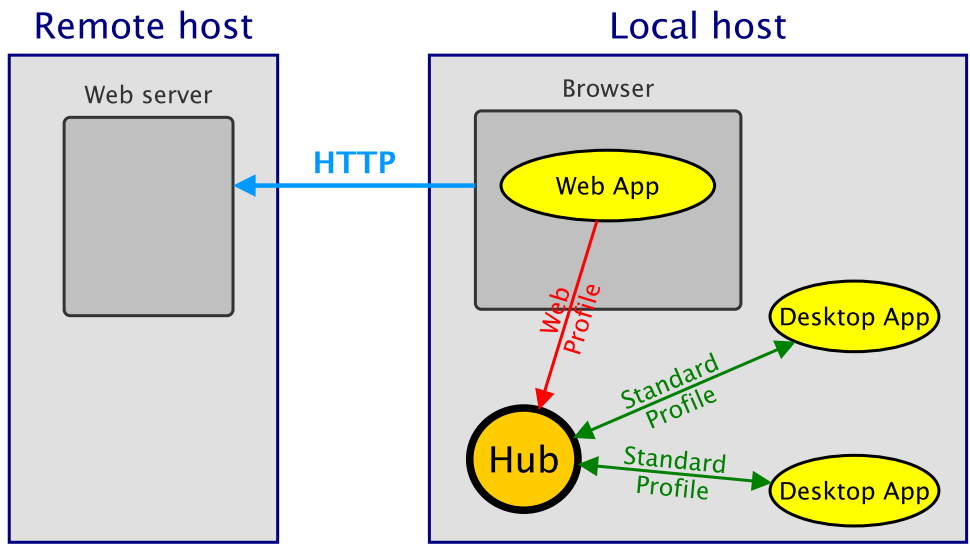}
\caption{Desktop communications using SAMP, including both desktop
         and browser-based SAMP clients.
         \label{P2-7:fig:sampcomms}}
\end{figure}

Web applications (typically HTML\,+\,JavaScript)
sometimes want to communicate with desktop
applications too, but browser sandboxing means that the
Standard Profile cannot be used,
so SAMP v1.3 \citep{2012ivoa.spec.1104T}
defined the {\em Web Profile\/},
which uses a well-known port, cross-origin workarounds and
message polling to provide the required functionality.
The general architecture of a SAMP desktop is illustrated in
Figure~\ref{P2-7:fig:sampcomms}.

Web SAMP is used in a number of web pages,
but in practice SAMP interactions from web pages
nearly all seem to follow the same pattern:
the result of some archive search contains a button like
``Send table via SAMP'' or ``Send FITS image via SAMP''.
This allows the user to take the result of a query
made on the web and insert it directly into a chosen SAMP-capable
desktop application such as TOPCAT or ds9 with one click.
This is a nice convenience, but it's really just saving
the user from having to save from the browser to disk
and then reload into the client application.
There seem to be very few Web SAMP applications that
offer interoperability functions beyond exchanging a
table or image;
the only one known at time of writing is the WorldWide Telescope
interface to the Chandra Source Catalog,
which can exchange sky pointing messages with other clients.

\section{The Problem with HTTPS}

HTTPS (secure HTTP) is HTTP layered over TLS (Transport Layer Security).
Alongside encryption, this transport layer enforces host authentication
by requiring the server to present to the client a trusted certificate.
Driven by security concerns, and possibly buzzword compliance,
data providers are increasingly replacing HTTP services with
HTTPS equivalents.

Although SAMP's Web Profile works well with HTTP-based web applications,
it cannot be made to work for web applications hosted on servers using the
HTTPS protocol.

The reasons for this are somewhat involved.
If the browser retrieves a web page from a remote host using HTTPS,
that page is only permitted to contact other URLs that are
also HTTPS, since accessing HTTP pages would potentially compromise
the integrity of content that HTTPS assures.
The SAMP Web Profile requires web applications
to communicate with the Hub
at the local URL {\tt http://localhost:21012/},
which is done
using the browser-provided {\tt XMLHttpRequest} API from JavaScript
or something similar.
If the web application has been downloaded from an HTTPS service,
browser sandboxing prevents it from doing that.
This is reported by the browser (if you know where to look)
with a message reporting that
{\em Mixed Active Content\/} has been blocked when attempting
to access the local hub HTTP server.

The obvious solution would be to run the local hub service
over HTTPS rather than HTTP.
Unfortunately, this is much harder than it sounds.
In order to serve HTTPS, the server must be able to present
a {\em trusted certificate},
that is one recognised by the browser.
For a well-known URL of the form
{\tt https://localhost[:<port>]/<path>}
the hub server would need a certificate
for the domain ``{\tt localhost}'';
but Certificate Authorities are not permitted
to issue certificates for that domain,
or for equivalent loopback addresses.
It would be possible to self-sign a certificate for this domain,
but then the browser would not recognise it
(in the absence of browser security reconfiguration,
which normal users don't want and shouldn't be encouraged to do).
Conceivably one could acquire a certificate from a CA for a domain
that DNS-resolves to 127.0.0.1;
that's more effort than the SAMP user can be expected to make,
but if done by the hub developer, then public/private key pairs
would have to be distributed with the hub, which feels very wrong.
In any case, where CAs have spotted this sort of thing in the past,
it seems they have generally revoked the certificate.
In principle it would be possible for the user to acquire a certificate
associated with the actual hostname of their machine and install it
into the hub,
but even apart from the work and expense required from the user,
there are still problems for the web app to determine what the
hostname actually is in order to contact the server.

In short, running an HTTPS service for access on a well-known local host
URL seems to be impossible.

\section{Is an HTTPS Profile Possible?}
\label{P2-7:sec:httpsprof}

A scheme for achieving SAMP communication between an HTTPS-hosted
web application (the SAMP client) and a local hub has been worked
out and prototyped, but it is far from elegant.

The basic idea is that since the client can't request HTTP,
and the Hub can't serve HTTPS, the two parties have to use
an intermediary for communications.
An HTTPS service known as the {\em Relay\/} is therefore
set up on an external machine, and both client and hub
use this to forward messages to each other.
The question remains of how the hub knows where and when to
contact the relay service.
This can be solved by abuse of a loophole in current browser security
policies which allows mixed {\em passive\/} content;
the client is permitted to request an image from the local hub
server over HTTP using the HTML {\tt IMG} tag, and
it uses the text of this image request URL to smuggle
a message about the relay location to the hub,
a manoevre we call the {\em Nudge}.

This scheme has been prototyped and is in experimental use at SSDC,
but it has several disadvantages.
First, all the communications between the client and hub,
both running on the local host, are routed via a potentially distant
external host, with impacts on latency, reliability and security.
Second, the scheme relies on the use of mixed passive content;
current browsers indicate this as a reduced security status
for instance with a ``warning padlock'' address bar icon,
and future browser policy may block such usages altogether
(such a change is the stated intention of the W3C
{\em Mixed Content\/} specification).
Finally, the whole scheme is complex to specify and implement,
with plenty of opportunities for implementation issues.

\section{What Next?}

We present for discussion three options for the future of Web SAMP
over HTTPS.

\paragraph{1. Standardise the HTTPS Profile}

It would be possible to issue a new version of the SAMP standard
incorporating an {\em HTTPS Profile\/} as outlined in
Section~\ref{P2-7:sec:httpsprof}.
This would involve drafting the additional text,
providing two independent implementations of the new functionality
(the existing prototype is in Java; probably a Python one would
also be required),
pushing the new version through the IVOA Recommendation process,
and ensuring that users are working with HTTPS-Profile-capable Hub
implementations by embedding the updated hub in popular SAMP clients.
Data providers adopting the new profile would need to deploy Relay
services alongside their existing web applications,
which makes the process of providing SAMP-capable web pages more complex.
The standardisation process in particular is time consuming
in terms of both effort and elapsed time,
especially for a technically complex enhancement like this.
There remain also some loose ends to tie up in the existing prototype
implementation.

If done, this would allow SAMP to work equally (though not perhaps
equally well) from HTTP and HTTPS; however it would require considerable
effort, not be in place for some while, and may stop working in the
future if browser security policies change.

\paragraph{2. Use a SAMP-capable helper application}

Since by far the most common use of Web SAMP is to ask
desktop applications to load a VOTable or FITS file,
we could get away with something much simpler than a full SAMP client.
One possibility is a simple helper application for use with browsers
that accepts a filename on the command line and forwards it to
running SAMP-capable desktop clients.
The user could either associate this helper with suitable MIME types,
or just choose it at the browser's {\em Open with...} prompt.
This low-tech solution requires no strange or questionable tricks,
and an example implementation {\tt sampload} has been implemented
and is available as part of the JSAMP library (version 1.3.6).

This solution is simple to implement and
requires no new standardisation activity;
unlike Web SAMP however, it doesn't just work with one click,
the user is required to download and install the helper application.

\paragraph{3. Do Nothing}

The idea of using SAMP from HTTPS pages could simply be abandoned.
Web SAMP will still work from HTTP pages, but not from the
increasing number of pages served using HTTPS.
Users of those services will just have to save files to local disk
and reload them into a suitable local application rather than
use a ``send to TOPCAT'' button or similar.
This is a loss of convenience, but since more complex uses of
Web SAMP such as exchanging coordinates between web and desktop
clients are uncommon, it won't have much impact on what users
are actually able to do.

\bibliography{P2-7}

% if we have space left, we might add a conference photograph here. Leave commented for now.
% \bookpartphoto[width=1.0\textwidth]{foobar.eps}{FooBar Photo (Photo: Any Photographer)}

\end{document}